\documentclass[aps,prb,twocolumn,floatfix,showpacs,citeautoscript,longbibliography,hyperlinks,superscriptaddress]{revtex4-1}

\usepackage{graphicx}
\usepackage{dcolumn}
\usepackage{bm}
\usepackage{bbold}
\usepackage{mathtools}
\usepackage{color}
\usepackage{transparent}
\usepackage[colorlinks=true,citecolor=blue,bookmarksdepth=2]{hyperref}
\usepackage{framed}

\DeclareMathOperator{\tr}{Tr}
\newcommand*{\citen}[1]{%
  \begingroup
    \romannumeral-`\x 
    \setcitestyle{numbers}%
    \cite{#1}%
  \endgroup
}

\begin{document}
\title{Photon counting statistics of a microwave cavity}
\author{Fredrik Brange}
\affiliation{Department of Physics and NanoLund, Lund University, Box 188, SE-221 00 Lund, Sweden}
\author{Paul Menczel}
\affiliation{Department of Applied Physics, Aalto University, 00076 Aalto, Finland}
\author{Christian Flindt}
\affiliation{Department of Applied Physics, Aalto University, 00076 Aalto, Finland}

\begin{abstract}
The development of microwave photon detectors is paving the way for a wide range of quantum technologies and fundamental discoveries involving single photons. Here, we investigate the photon emission from a microwave cavity and find that distribution of photon waiting times contains information about few-photon processes, which cannot easily be extracted from standard correlation measurements. The factorial cumulants of the photon counting statistics are positive at all times, which may be intimately linked with the bosonic quantum nature of the photons. We obtain a simple expression for the rare fluctuations of the photon current, which is helpful in understanding earlier results on heat transport statistics and measurements of work distributions. Under non-equilibrium conditions, where a small temperature gradient drives a heat current through the cavity, we formulate a fluctuation-dissipation relation for the heat noise spectra. Our work suggests a number of experiments for the near future, and it offers theoretical questions for further investigation.
\end{abstract}

\maketitle
\section{Introduction}
The development of quantum technologies relies on the ability to control, transmit, and detect single quanta of light, heat, and charge~\cite{zagoskin:2011}. Much effort has thus been devoted to the manipulation of individual photons \cite{Lang2011,PhysRevLett.112.116802}, phonons \cite{Clerk2010,NatureLetter520}, and electrons \cite{Splettstoesser2017} at the nano-scale. Electrons \cite{Feve2007,Bocquillon2013,Dubois2013,Jullien2014} and photons \cite{Aharonovich2016} can be emitted on demand and in some cases detected with single-particle resolution. In one approach, single electrons are captured in a quantum dot, whose charge state is read out using a capacitively coupled conductor~\cite{GUSTAVSSON2009191}. Photons, by contrast, are uncharged with energies in nanoscale systems that can be very small (in the microwave range), requiring highly sensitive detectors~\cite{Gu2017}.

Recently, it has been suggested that microwave photons may be detected in a calorimetric approach~\cite{Pekola2013,Gasparinetti2015,Brange2018}. A resistive environment is monitored in real-time using ultrasensitive thermometry with dips and peaks in the temperature corresponding to the emission or absorption of single photons. In another proposal, microwave photons are detected using Josephson junctions \cite{Chen2011,Walsh2017}. Very recently, a quantum non-demolition detector for propagating microwave photons was realized~\cite{Besse2018}. Such single-photon detectors are paving the way for a wide range of applications within quantum thermodynamics \cite{Vinjanampathy2016}, feedback control~\cite{wiseman:2009}, and quantum information processing~\cite{nielsen:2011}. Moreover, they may help address fundamental questions regarding heat transport, entropy production, and fluctuation relations at the nanoscale \cite{Esposito2009}.

\begin{figure}
  \centering{
\includegraphics[width=0.46\textwidth]{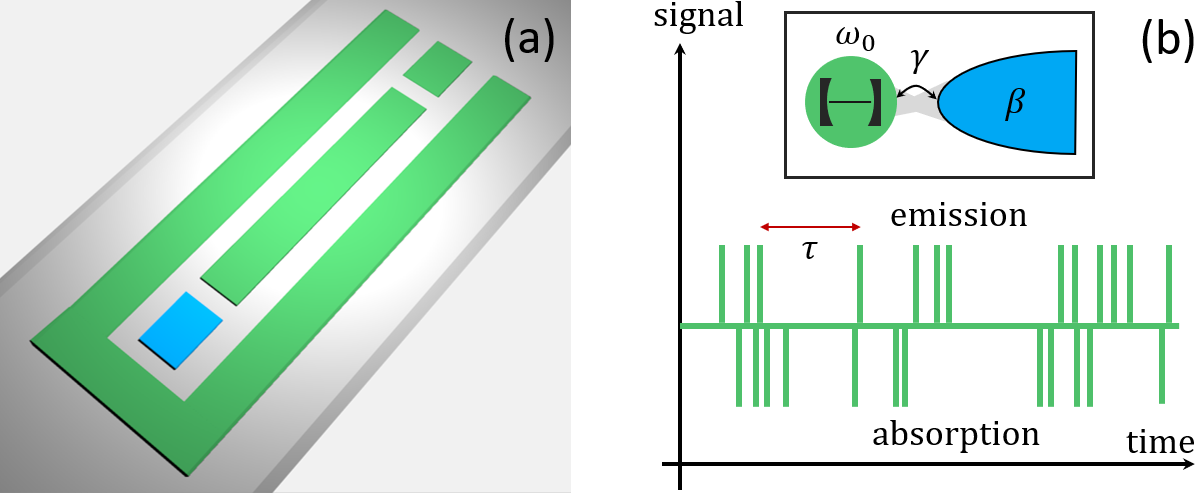}}
  \caption{Photon emission from a microwave cavity. (a) Photons are transmitted between a microwave cavity (in green) and an external heat bath (in blue). (b) A possible time trace of emission and absorption events measured by a single-photon detector. The waiting time between photon emissions is denoted by $\tau$.
  The setup is shown schematically in the inset. The cavity with frequency $\omega_0$ is coupled at the rate $\gamma$ to an external heat bath at the inverse temperature $\beta$. }
\label{Microwave cavity}
\end{figure}

In this work, we investigate the photon counting statistics of a microwave cavity at the single-particle level~\cite{Liu2015,Mi2017,Mi2018,Landig2018}, see Fig.~\ref{Microwave cavity}. The problem is simple to formulate, yet, surprisingly rich in physics. By combining a generating function technique with the method of characteristics, we obtain a full analytic solution for the photon counting statistics on all relevant timescales. The short-time physics can be characterized by the distribution of photon waiting times~\cite{Vyas1989,PhysRevA.39.1200,Zhang2018,PhysRevLett.108.186806,PhysRevB.90.205429,PhysRevB.91.195420}, which contains information about few-photon processes which cannot easily be extracted from standard correlation measurements. The factorial cumulants of the counting statistics~\cite{PhysRevLett.81.1829,PhysRevLett.86.700,PhysRevLett.81.1829,PhysRevB.83.075432,Kambly2013,PhysRevB.92.155413,Kleinherbers2018,Komijani2013} are positive at all times, and we conjecture that this behavior is linked with the bosonic quantum nature of the photons. At long times, we find a simple expression for the rare fluctuations of the photon current which may explain earlier results on heat transport statistics~\cite{PhysRevLett.99.180601} and measurements of work distributions \cite{0295-5075-89-6-60003}. Finally, we consider a non-equilibrium situation, where a temperature gradient drives a heat current through the cavity. Here, we obtain fluctuation-dissipation theorems in the linear and weakly non-linear regimes, and we formulate a relation between the heat noise spectra and the response of the system to small perturbations of the cavity frequency.

\section{Microwave cavity}

We consider the photon emission from a microwave cavity with the Hamiltonian $\hat{H} = \hbar \omega_0 \left(\hat{a}^\dagger \hat{a}+\frac{1}{2}\right)$, where $\hat{a}^\dagger$ ($\hat{a}$) creates (annihilates) photons with frequency $\omega_0$. The density matrix of the cavity $\hat{\rho}(t)$ evolves according to the Lindblad equation~\cite{breuer2007theory}
\begin{equation}
\frac{d\hat{\rho}}{dt} = \mathcal{L}\hat{\rho} = -\frac{i}{\hbar}[\hat{H},\hat{\rho}]+\gamma\left([\bar{n}+1]\mathcal{D}[\hat{a}]\hat{\rho}+\bar{n}\mathcal{D}[\hat{a}^\dagger]\hat{\rho}\right),
\label{Lindblad equation}
\end{equation}
where $\bar{n} = 1/(e^{\beta \hbar\omega_0}-1)$ is the average occupation of the cavity in equilibrium at the inverse temperature $\beta=1/(k_BT)$, and $\gamma$ governs the photon emission and absorption rates. The Liouvillian $\mathcal{L}$ captures both the unitary evolution described by $\hat{H}$ and the incoherent dynamics given by the dissipators, $\mathcal{D}[\hat{a}]\hat{\rho} \equiv \hat{a}\hat{\rho}\hat{a}^\dagger-\frac{1}{2}\{\hat{a}^\dagger{\hat{a}},\hat{\rho}\}$. To be specific, we formulate our problem in terms of a microwave cavity~\cite{Liu2015,Mi2017,Mi2018,Landig2018}, however,  our findings below are clearly valid for any other bosonic degree of freedom that can be treated as a dissipative quantum harmonic oscillator, for instance, a nano-mechanical resonator~\cite{NatureLetter520}. Moreover, the heat bath can be either bosonic or fermionic (see App.~\ref{Appendix A}), as for example an electronic reservoir, where the emission and absorption of single photons give rise to dips and peaks in the temperature, which can be measured using ultrasensitive thermometry~\cite{Pekola2013,Gasparinetti2015,Brange2018}.

\section{Photon counting statistics}

To investigate the photon counting statistics, we unravel the Lindblad equation with respect to the number of photons $m$ emitted during the time span $[0,t]$ \cite{RevModPhys.70.101}. Hence, we resolve the density matrix as $\hat{\rho}(t)=\sum_m \hat{\rho}(m,t)$, from which we obtain the photon counting statistics, $P(m,t)=\mathrm{Tr}\{\hat{\rho}(m,t)\}$. The density matrices evolve as $\frac{d}{dt}\hat{\rho}(m,t) =(\mathcal{L}-\mathcal{J}_e) \hat{\rho}(m,t) +\mathcal{J}_e\hat{\rho}(m-1,t)$, where $\mathcal{J}_{e}\hat{\rho}= \gamma(\bar{n}+1)\hat{a}\hat{\rho}\hat{a}^\dagger$ is the superoperator for the photon emission current. The equations of motion do not couple populations of the density matrices to the coherences, however, the populations are mutually coupled. To decouple the system of equations, we introduce the generating function $\mathcal{G}(s,q,t) \equiv \sum_{n,m} \langle n|\hat{\rho}(m,t)|n\rangle e^{ms+nq}$, where $s$ and $q$ are conjugate variables to the number of emitted photons $m$ and the cavity occupation number $n$, respectively. The generating function obeys the partial differential equation (see App.~\ref{Appendix A})
\begin{equation}
\partial_t \mathcal{G}(s,q,t) = [f(s,q)+g(q)]\partial_q \mathcal{G}(s,q,t) + g(q)\mathcal{G}(s,q,t),
\label{Partial differential equation}
\end{equation}
with $f(s,q) = \gamma(\bar{n}+1)(e^{s-q}-1)$ and $g(q) = \gamma \bar{n}(e^q-1)$. Remarkably, the differential equation can be solved analytically using the method of characteristics \cite{RenardyRogers}. The generating function contains statistical information both about the number of photons in the cavity~\cite{Arnoldus:96,Clerk2007,Clerk2011,Hofer2016} and the number of photons that have been emitted~\cite{Bedard1966,Tornau1973,Mehta1975}.  Here, we focus on the photon emission statistics with the moment generating function $\mathcal{M}(s,t)\equiv\sum_{m} P(m,t) e^{ms}=\mathcal{G}(s,0,t)$. In thermal equilibrium, we find (see App.~\ref{sec:mgf})
\begin{equation}
\mathcal M(s,t) = \frac{2\xi\, e^{\gamma t / 2}}{2\xi \cosh\!\big[ \frac{\xi\gamma t}{2}\big] + (1+\xi^2) \sinh\!\big[ \frac{\xi\gamma t}{2}\big]},
\label{MGF}
\end{equation}
with $\xi = \sqrt{1-4\bar{n}(1+\bar{n})(e^s-1)}$. This expression holds on all timescales, where Eq.~\eqref{Lindblad equation} is valid~\cite{Note1}, and it is important for our further analysis of the photon emission statistics. With $\gamma$ fixing the timescale, we are left with a single dimensionless parameter, namely the mean occupation number~$\bar{n}$, controlled by the temperature $T$.

\section{Waiting time distribution}

We first analyze the waiting time $\tau$ between photon emissions~\cite{Vyas1989,PhysRevA.39.1200,Zhang2018}. Recently, waiting time distributions have been measured both for photon emission \cite{PhysRevLett.112.116802} and electron tunneling \cite{PhysRevApplied.8.034019}.  The waiting time distribution can be obtained as $\mathcal{W}(\tau) = \langle \tau \rangle \partial_\tau^2 \Pi(\tau)$, where $\langle \tau \rangle$ is the mean waiting time and $\Pi(\tau)$ is the probability that no photons are emitted in a time span of duration $\tau$ \cite{PhysRevLett.108.186806,PhysRevB.90.205429}. Physically, the time-derivatives correspond to a photon emission at the beginning and the end of the time interval \cite{PhysRevB.91.195420}. From the definition of the moment generating function, we have $\Pi(\tau) = \mathcal{M}(-\infty,\tau)$ and then obtain (see App.~\ref{Appendix C})
\begin{equation}
	\mathcal W(\tau) = \Gamma\gamma\bar\gamma \frac{\gamma + 6\Gamma + (\gamma+2\Gamma) \cosh[\bar\gamma t] + \bar\gamma \sinh[\bar\gamma t]}{\left( \bar\gamma \cosh\!\big[ \frac{\bar\gamma t}{2} \big] + (\gamma + 2\Gamma) \sinh\!\big[ \frac{\bar\gamma t}{2} \big] \right)^3} e^{\frac{\gamma t}{2}}.
\label{WTD}
\end{equation}
Here, we have used that the average emission rate is $\langle J_e\rangle=\gamma \bar{n}(1+\bar{n}) \equiv \Gamma$, and we have defined $\bar\gamma \equiv \gamma(1+2\bar{n})$.

\begin{figure*}
  \centering
  \includegraphics[scale=0.56]{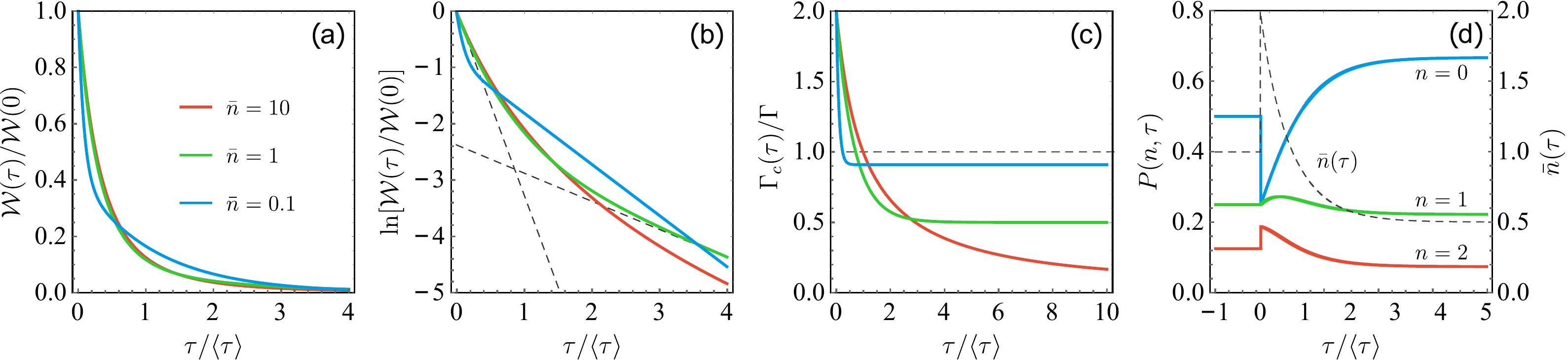}
  \caption{Photon waiting times. (a) Distribution of waiting times for different average occupations of the cavity, $\bar{n}=0.1,1,10$. (b) The fast and slow decay rates are clearly visible on a logarithmic scale with the dashed lines corresponding to $\bar{n}=1$. (c) Conditional emission rate given that the last emission occurred at the time $\tau=0$. The dashed line corresponds to a Poisson process. (d) Probability of having $n$ photons (left axis) and mean number of photons (right axis) in the cavity given that the last emission occurred at the time $\tau=0$. For $\tau <0$, the cavity is populated according to a Boltzmann distribution with $\bar{n}=1$.}
\label{WTDs}
\end{figure*}

Figure~\ref{WTDs}~(a) shows waiting time distributions for different temperatures. The distributions start off at a finite value, $\mathcal{W}(0) = 2\Gamma$, and then decay monotonically to zero at long times. This behavior should be contrasted with that of noninteracting fermions, for which the distributions are typically suppressed at short times due to the Pauli principle \cite{PhysRevLett.108.186806,PhysRevB.90.205429}. Similarly to the recent experiments \cite{PhysRevLett.112.116802,PhysRevApplied.8.034019}, the waiting time distributions are double-exponential. For short times, $\Gamma\tau \ll 1$, we have $\mathcal{W}(\tau) \simeq 2\Gamma\exp{(-\gamma[6\bar{n}(\bar{n}+1)+1]\tau/2)}$, showing that the fast decay rate increases quadratically with the mean occupation $\bar{n}$, and not just linearly as one might expect. At long times, $\Gamma\tau \gg 1$, we have $\mathcal{W}(\tau)\simeq\frac{4\Gamma\gamma\bar\gamma}{(\gamma+\bar\gamma+2\Gamma)^2}\exp{(-\gamma\bar{n}\tau)}$ with the slow decay rate given by~$\gamma\bar{n}$. Figure \ref{WTDs}~(b) illustrates the cross-over between these limiting behaviors.

The increased decay rate at short times is a signature of photon bunching. This phenomenon is illustrated in Fig.~\ref{WTDs}~(c), showing the conditional emission rate, $\Gamma_c(\tau) = \mathcal{W}(\tau)\big/\int_\tau^\infty \mathcal{W}(u) du$, at the time $\tau$ after the last photon emission. Due to the photon bunching, the rate is enhanced at short times and suppressed at long times. The bunching also affects the number of photons in the cavity at the time~$\tau$ after the last photon emission, see Fig.~\ref{WTDs}~(d). An application of Bayes' theorem shows that the expected number of photons in the cavity increases by a factor of two directly after an emission event (see App.~\ref{Appendix C}). At longer times, with no subsequent emissions, it is increasingly likely that the cavity is empty, and it eventually reaches a Boltzmann distribution, albeit with an average photon number $\bar{n}/(1+\bar{n})$ suppressed below one.

\section{Correlation function}

A different perspective on the short-time physics is provided by the $g^{(2)}$-function \cite{Ottl2005,Lang2011,PhysRevLett.112.116802,NatureLetter520}. The $g^{(2)}$-function is proportional to the probability that a  photon is emitted at the time $\tau$, given that a photon was emitted at the time $\tau=0$. Unlike the waiting time distribution, other photon emissions may have occurred during this time span. The correlation function can be obtained from  Eq.~(\ref{MGF}), and we find (see App.~\ref{sec:g2})
\begin{equation}
g^{(2)}(\tau) = 1+e^{-\gamma |\tau|}.
\label{SDC}
\end{equation}
This is the $g^{(2)}$-function for chaotic thermal light as well as for other non-interacting bosons, for example, thermal phonons as shown in recent experiments~\cite{NatureLetter520}. Equations (\ref{WTD}) and (\ref{SDC}) are important for the recurring discussion about possible connections between the waiting time distribution and the $g^{(2)}$-function. For renewal processes, where consecutive waiting times are uncorrelated, the two functions are related in Laplace space as $g^{(2)}(s)\, \langle J_e \rangle = \mathcal{W}(s) / [1\!-\!\mathcal{W}(s)]$  \cite{PhysRevB.85.165417,PhysRevB.90.205429,PhysRevB.91.195420,PhysRevB.85.165417}. This relation does not hold for our cavity, since it does not return to the same state after each emission. Moreover, unlike the $g^{(2)}$-function, the waiting time distribution depends on temperature, showing that the two are not equivalent.

\section{Factorial cumulants}

To investigate the transition from short to long observation times, we consider the factorial cumulants of the photon counting statistics~\cite{PhysRevLett.81.1829,PhysRevLett.86.700,PhysRevLett.81.1829,PhysRevB.83.075432,Kambly2013,PhysRevB.92.155413,Kleinherbers2018,Komijani2013}. The factorial cumulants are defined as $\langle\! \langle m^k\rangle\!\rangle_F = \langle\!\langle m(m-1)...(m-k+1)\rangle\!\rangle$, where $\langle \! \langle m^k \rangle\!\rangle = \partial_s^k \ln\mathcal{M}(s,t)|_{s=0}$ are the ordinary cumulants of order $k$. The counting statistics of noninteracting electrons in a two-terminal setup is always generalized binomial \cite{PhysRevB.78.165330,PhysRevLett.100.086602,PhysRevB.79.205315}, and the sign of the factorial cumulants alternates with the order~$k$~\cite{PhysRevB.83.075432,Kambly2013,PhysRevB.92.155413,Kleinherbers2018}. By contrast, for the photon cavity we find
\begin{equation}
\begin{split}
\langle \! \langle m \rangle \!\rangle_F &=  \gamma t \bar{n}(1+\bar{n}), \\
\langle \! \langle m^2 \rangle \!\rangle_F &=  2\gamma t \left[1+\frac{e^{-\gamma t}-1}{\gamma t}\right]\bar{n}^2(1+\bar{n})^2,\\
\langle \! \langle m^3 \rangle \!\rangle_F &=12\gamma t \left[1+e^{-\gamma t}+2\frac{e^{-\gamma t}-1}{\gamma t} \right] \bar{n}^3(1+\bar{n})^3,
\end{split}
\end{equation}
with similar expressions for the higher factorial cumulants, which are positive. These results suggest that the quantum statistics of the particles, being bosons or fermions, is intimately linked with the sign of the factorial cumulants, consistently with earlier works on photon counting statistics~\cite{PhysRevLett.81.1829,PhysRevLett.86.700}. At long observation times, we have $\langle\!\langle m^k\rangle\!\rangle_F \propto \gamma t \bar{n}^k(1+\bar{n})^k$, showing that the photon counting statistics is nearly Poissonian at low temperatures, where only the first factorial cumulant is non-zero.

\section{Long-time statistics}

To complete the discussion of the long-time limit, we analyze the large-deviation statistics of the photon emission current~\cite{TOUCHETTE20091}. To this end, we evaluate the counting statistics $P(J_e,t) = \frac{1}{2\pi i}\int_{-i\pi}^{i\pi} ds\, e^{t\left[\Theta(s)-sJ_e\right]}$ in the long-time limit, where $\Theta(s)=\lim_{t\rightarrow\infty} \ln\left[\mathcal{M}(s,t)\right]/t$ is the cumulant generating function for the photon emission current~$J_e=m/t$,
\begin{equation}
\Theta(s)  =\frac{\gamma}{2}\left(1-\sqrt{1-4(e^s-1)\bar{n}(1+\bar{n})}\right).
\label{eq:CGFemis}
\end{equation}
The large-deviation statistics of the emission current can be evaluated in a saddle-point approximation,
\begin{equation}
\frac{\ln[P(J_e,t)]}{t} \simeq \Theta(s_\mathrm{o})-s_\mathrm{o}J_e,
\label{Saddle-point approximation}
\end{equation}
where $s_\mathrm{o}=s_\mathrm{o}(J_e)$ solves the saddle-point equation $\Theta'(s_\mathrm{o})= J_e$. Figure~\ref{Long-time statistics}~(a) shows the large-deviation statistics for different temperatures. With increasing temperature, the distributions become strongly non-Poissonian and large emission currents are more likely. For large currents,  the saddle-point $s_\mathrm{o}$ must be close to the square-root singularity of $\Theta(s)$ at $s=s_c$, where $\Theta(s_c)=\gamma/2$ and the derivative $\Theta'(s_c)$ diverges. With $s_\mathrm{o}\simeq s_c=2\ln\!\left[\cosh\left(\beta\hbar\omega_0/2\right)\right]\simeq \beta\hbar\omega_0$ for $\beta\hbar\omega_0\gg1$, the large-deviation statistics becomes (see App.~\ref{Appendix E})
\begin{equation}
\frac{\ln[P(J_e,t)]}{t}  \simeq \gamma/2-\beta\hbar\omega_0J_e,\,\, J_e \gg\gamma.
\label{eq:LDFapprox}
\end{equation}
This expression agrees well with the exact results in Fig.~\ref{Long-time statistics}~(a). As we discuss below, it provides an analytic understanding of the linear dependence on the heat current and the inverse temperature observed in numerical calculations of the large-deviation statistics in phononic heat transport \cite{PhysRevLett.99.180601}. A similar reasoning might also be helpful in understanding the tails of the work distributions measured for a micro-cantilever \cite{0295-5075-89-6-60003}.

\section{Heat transport}

\begin{figure}
  \centering
  \includegraphics[scale=0.60]{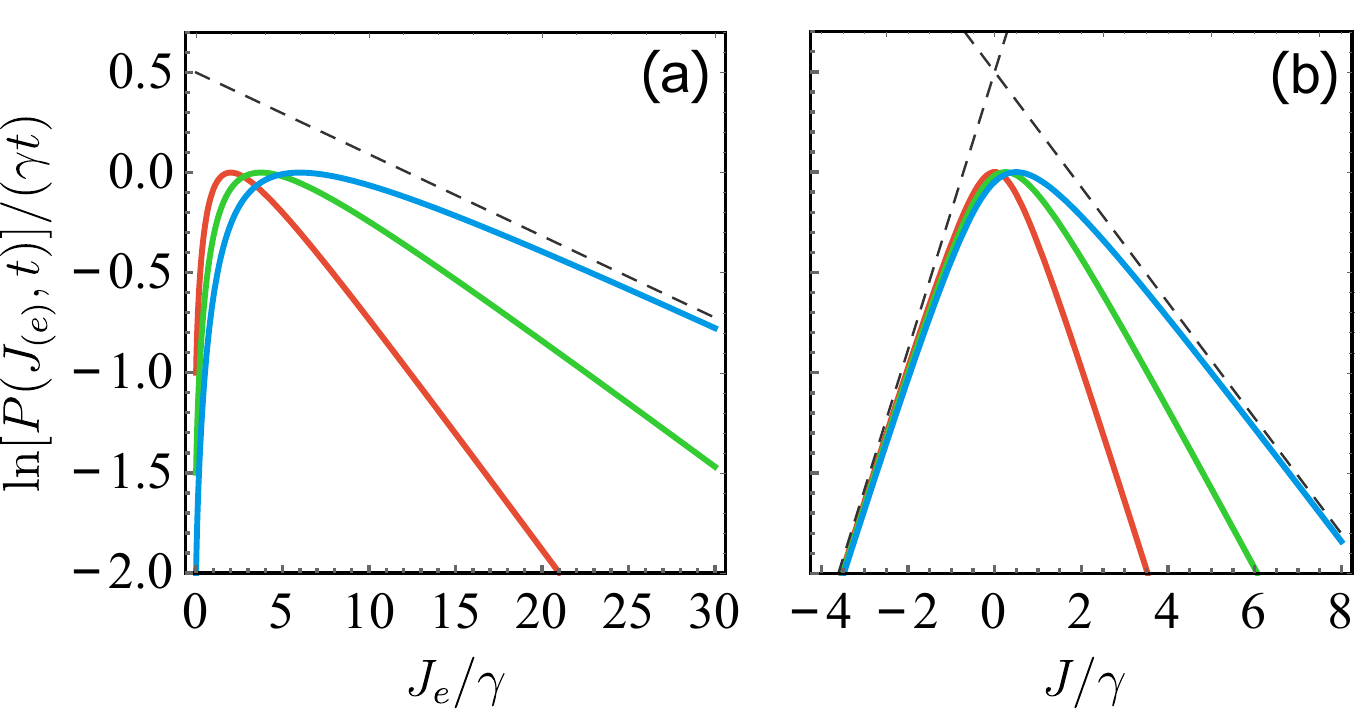}
  \caption{Large-deviation statistics of the photon current. (a)  Analytic results for the distribution of the photon emission current $J_e$ from a cavity coupled to a single reservoir with $\bar{n} = 1$ (red), $1.5$ (green) and $2$ (blue). The dashed line is based on the branch-point of the cumulant generating function in Eq.~(\ref{eq:CGFemis}), and it is given by Eq.~(\ref{eq:LDFapprox}) at low temperatures. (b) Distribution of the current $J$ running via the cavity between a hot and a cold reservoir with $\bar{n}_c=1$, $\bar{n}_h=1$ (red), $\bar{n}_c=1$, $\bar{n}_h=2$ (green) and $\bar{n}_c=1$, $\bar{n}_h=3$ (blue), and we have defined $\gamma\equiv\gamma_h+\gamma_c$. The dashed lines are approximations based on the branch-points of the cumulant generating function in Eq.~(\ref{Long-time CGF net current}). At low temperatures, the left line is of the form $\propto\beta_c\hbar\omega_0J$, while the right one is given by $\propto-\beta_h\hbar\omega_0J$.}
\label{Long-time statistics}
\end{figure}

Our analysis can be extended to setups with the cavity coupled to several reservoirs kept at different temperatures, thus providing an interesting opportunity to investigate the heat flow through the cavity in a non-equilibrium situation~\cite{PhysRevLett.112.076803,Tang2018,Denzler2018,Salazar2018}. Similar to Eq.~\eqref{MGF}, we can evaluate the moment generating function at finite times for the transfer of photons between the cavity and each reservoir~\cite{Note2}. Here we are particularly interested in the long-time statistics of the photon current $J$ running via the cavity from a hot to a cold reservoir. For the net photon current, the cumulant generating functions reads
\begin{equation}
\Theta(s) = \frac{\gamma_c+\gamma_h}{2}\left(1-\sqrt{1-4\frac{\gamma_c\gamma_h}{(\gamma_c+\gamma_h)^2}\kappa(s)}\right),
\label{Long-time CGF net current}
\end{equation}
with $\kappa(s)\equiv (e^s-1)(1+\bar{n}_c)\bar{n}_h+(e^{-s}-1)\bar{n}_c(1+\bar{n}_h)$, where $\bar{n}_{h(c)}$ is the Bose-Einstein distribution of the hot (cold) bath at the photon frequency $\omega_0$, and $\gamma_{h(c)}$ is the coupling strength (see App.~\ref{sec:generalization}). This expression also holds for the heat exchange between two resistors connected via a narrow transmission profile \cite{PhysRevB.92.085412}. Again, we can evaluate the large-deviation statistics by analytically solving the saddle-point equation. The cumulant generating function has square-root singularities both for positive and negative values of $s$, which determine the linear parts of the large-deviation function for large (positive or negative) photon currents as illustrated in Fig.~\ref{Long-time statistics}~(b). These results resemble the numerical findings of Ref.~\onlinecite{PhysRevLett.99.180601}.

\section{Fluctuation relations}

It is interesting to understand the properties of the heat current fluctuations. The cumulant generating function fulfills the symmetry $\Theta(s) = \Theta(-s-\sigma)$, where $\sigma=\hbar\omega_0(\beta_c-\beta_h)$ determines the entropy increase per transferred photon. This symmetry immediately implies the fluctuation relation \cite{Gallavotti1995,PhysRevLett.74.2694} (see App.~\ref{Appendix G})
\begin{equation}
\frac{1}{t}\ln\left[\frac{P(J,t)}{P(-J,t)} \right] = \sigma J ,
\label{FR}
\end{equation}
which connects the probabilities to observe photon currents $J$ of opposite signs, also far from equilibrium with large temperature differences. Close to equilibrium, we may expand the mean heat current $\langle J_Q\rangle \equiv \hbar\omega_0\langle J\rangle\simeq G_Q^{(1)}\Delta T +G_Q^{(2)}\Delta T^2/2$ and the noise $S_Q=\langle\!\langle J_Q^2 \rangle\!\rangle \simeq S_Q^{(\mathrm{eq})}+ S^{(1)}_Q\Delta T$ in the temperature difference $\Delta T$. From the symmetry of the generating function, we then obtain the fluctuation-dissipation theorem for heat currents, $S_Q^{(\mathrm{eq})} = 2k_BT^2G_Q^{(1)}$, relating the equilibrium noise to the linear thermal conductance~\cite{PhysRevLett.99.180601,PhysRevLett.104.076801}. Moreover, we find the relation $S^{(1)}_Q=k_BT^2G_Q^{(2)}$ between the noise susceptibility and the second-order response coefficient of the heat current in the weakly non-linear regime (see App.~\ref{Appendix H}).

\section{Noise power spectrum}

Finally, we turn to the noise spectra of the heat currents. The finite-frequency noise can be obtained from the moment generating function at finite times using MacDonald's formula~\cite{0034-4885-12-1-304,FLINDT2005411,PhysRevB.75.045340}. In equilibrium, the auto-correlation functions read (see App.~\ref{Appendix I})
\begin{equation}
S_Q^{c,h}(\omega)=S_Q^{(\mathrm{eq})}\left(1+\frac{\gamma_{c,h}}{\gamma_{h,c}}\frac{\omega^2}{(\gamma_c+\gamma_h)^2+\omega^2}\right),
\label{Eq12}
\end{equation}
while for the real-part of the cross-correlator, we find
\begin{equation}
\mathrm{Re}[S_Q^{ch}(\omega)]=S_Q^{(\mathrm{eq})}\left(-1+\frac{\omega^2}{(\gamma_c+\gamma_h)^2+\omega^2}\right).
\label{Eq13}
\end{equation}
We see that $S_Q^{c}(0)=S_Q^{h}(0)=-\mathrm{Re}[S_Q^{ch}(0)]=S_Q^{(\mathrm{eq})}$, since there is no accumulation of photons in the cavity at low frequencies. Generally, we do not expect simple fluctuation-dissipation theorems for the individual heat currents at finite frequencies \cite{PhysRevLett.104.220601}. On the other hand, using the continuity equation $\dot{U}(t)=-[J_Q^c(t)+J_Q^h(t)]$ for the cavity energy and the outgoing heat currents, we can write the energy fluctuations as $\omega^2S_U(\omega)=S_Q^{c}(\omega)+S_Q^{h}(\omega) +2\mathrm{Re}[S_Q^{ch}(\omega)]$. Now, applying a weak perturbation $\hat{H}'(t)=\mathcal{K}(t)\hat{H}$, the change of the cavity energy $\langle \Delta U \rangle(\omega)=\chi(\omega)\mathcal{K}(\omega)$ in the Fourier domain can be expressed in terms of the susceptibility $\chi(\omega)$ in response to the force $\mathcal{K}(\omega)$ \cite{0034-4885-29-1-306}. We then arrive at the  fluctuation-dissipation theorem, $S_U(\omega)=2k_BT\, \mathrm{Im}[\chi(\omega)]/\omega$, which is valid for frequencies below the temperature, $\hbar\omega\ll k_BT$. Combining these expressions brings us to the relation
\begin{equation}
S_Q^{c}(\omega)+S_Q^{h}(\omega) +2\mathrm{Re}[S_Q^{ch}(\omega)]=2k_BT\omega\, \mathrm{Im}[\chi(\omega)],
\label{Eq14}
\end{equation}
between the sum of the noise spectra and the response of the system to small perturbations of the cavity frequency. We expect this relation to hold for many systems, where external reservoirs exchange heat via a central region.

\section{Conclusions}

We have fully determined the photon counting statistics of a quantum harmonic oscillator with dissipative Lindblad dynamics. To be specific, we have formulated our finding in terms of a microwave cavity, although our general results are valid for any quantum harmonic oscillator. The short-time physics can be characterized by the distribution of photon waiting times, which contains information about few-photon processes that cannot easily be extracted from standard correlation measurements. The factorial cumulants are positive at all times, unlike the case of noninteracting electrons for which the sign alternates with the order. This finding indicates that the quantum statistics of the particles, being bosons or fermions, determines the sign of the factorial cumulants. We have obtained a simple expression for the large-deviation statistics of the photon current, which may explain earlier results on heat transport fluctuations and measurements of work distributions. Finally, we have generalized our problem to a non-equilibrium situation, in which a temperature gradient drives a heat current through the cavity. In this case, we have derived fluctuation-dissipation theorems in the linear and weakly non-linear regimes and formulated a relation between the heat noise spectra and the response of the system to small perturbations of the cavity frequency. These predictions may be tested in future experiments with single-photon detectors or calorimetric measurements of heat currents.

\begin{acknowledgments}
We thank K.~Brandner, A.~A.~Clerk, F.~Hassler, V.~F.~Maisi, P.~P.~Potts, and P.~Samuelsson for useful discussions. The work was supported by the Swedish Research Council and the Academy of Finland (projects No. 308515 and 312299).
\end{acknowledgments}

\begin{appendix}
\section{From the Lindblad equation [Eq.~\texorpdfstring{\eqref{Lindblad equation}}{(\ref*{Lindblad equation})}] to the partial differential equation [Eq.~\texorpdfstring{\eqref{Partial differential equation}}{(\ref*{Partial differential equation})}]}\label{Appendix A}

We here derive the partial differential equation in Eq.~\eqref{Partial differential equation} from the Lindblad equation [Eq.~\eqref{Lindblad equation}] for a single light mode, with resonance frequency $\omega_0$, coupled to a single heat bath at temperature $T$ (we consider multiple baths in App. \ref{sec:generalization}).
The light mode is modeled as a quantum harmonic oscillator with the Hamiltonian
\begin{equation}
	\hat{H} = \hbar \omega_0 \Big( \hat{a}^\dagger \hat{a} +\frac{1}{2} \Big),
\end{equation}
where $\hbar$ is the reduced Planck constant and $\hat{a}$ ($\hat{a}^\dagger$) is the anniliation (creation) operator of the oscillator. Taking the coupling to the heat bath into account, the time evolution of the reduced density matrix $\hat{\rho}$ of the cavity is given by the Lindblad master equation \cite{Lindblad1976,GoriniJMathPhys1976,breuer2007theory}
\begin{align} \label{eq:lindblad}
\nonumber
	\frac{d \hat{\rho}}{d t} = -\frac{i}{\hbar}\, [ \hat{H},\hat{\rho}] &+ \gamma (\bar n+1)\, \Big( \hat{a}\hat{\rho} \hat{a}^\dagger-\frac{1}{2}\{\hat{a}^\dagger \hat{a},\hat{\rho}\}\Big)\\
&+ \gamma \bar n\, \Big(\hat{a}^\dagger\hat{\rho} \hat{a}-\frac{1}{2}\{\hat{a} \hat{a}^\dagger,\hat{\rho}\}\Big),
\end{align}
which is the same as Eq.~\eqref{Lindblad equation} in the main text. Here, $\gamma$ is a reference rate of relaxations and excitations in the system induced by the reservoir and
\begin{equation}
\bar n \equiv \frac{1}{e^{\beta \hbar \omega_0}-1}
\end{equation}
is the average occupation of the light mode in equilibrium at the inverse temperature $\beta \equiv 1/(k_B T)$.

The Lindblad equation \eqref{eq:lindblad} is not dependent on the microscopic details of the heat bath and can describe both bosonic and fermionic heat baths. To see this, we may rewrite the emission and absorption rates as
\begin{equation}
\gamma(\bar{n}+1) = \gamma \int \frac{dE}{\hbar\omega_0} f(E)[1-f(E+\hbar \omega_0)]
\end{equation}
and
\begin{equation}
\gamma\bar{n} = \gamma \int \frac{dE}{\hbar\omega_0} f(E)[1-f(E-\hbar\omega_0)]
\end{equation}
using the definitions of the Bose-Einstein and Fermi-Dirac distributions, $\bar{n}$ and $f(E)=1/(e^{\beta E}+1)$. We may think of the left-hand side of these equations as corresponding to a bosonic bath, such as the thermal background radiation, with $\bar n$ being the average number of bosons in the reservoir with energy $\hbar\omega_0$. Similarly, we may think of the right-hand sides in terms of  a fermionic bath, such as the Fermi sea of electrons in a nanoscale conductor. In this case, the emission of a photon from the cavity with energy $\hbar\omega_0$ is associated with the excitation of an electron with energy $E$ to a higher-lying state with energy $E+\hbar\omega_0$. The absorption of a photon is in a similar manner associated with an electron relaxing from energy $E$ to a lower-lying state with energy $E-\hbar\omega_0$. The electronic processes take place close to the Fermi level, $E_F=0$, where the electronic density of states is approximately constant, $g(E)\propto 1/(\hbar\omega_0)$.

\subsubsection*{Unraveling the master equation}

To keep track of the number $m$ of photons emitted into the heat bath, we introduce the $m$-resolved density matrices $\hat{\rho}(m,t)$, so that $P(m,t) = \tr \hat{\rho}(m,t)$ is the probability of having emitted $m$ photons to the heat bath.
They satisfy the unraveled Lindblad equation \cite{RevModPhys.70.101}
\begin{align}
\nonumber
	\frac{d \hat{\rho}(m,t)}{d t} &= -\frac{i}{\hbar}\, [ \hat{H},\hat{\rho}(m,t)] \\
\nonumber
&+ \gamma (\bar n+1)\, \Big(\hat{a}\hat{\rho}(m\!-\!1,t) \hat{a}^\dagger-\frac{1}{2}\{\hat{a}^\dagger \hat{a},\hat{\rho}(m,t)\}\Big) \\
&+ \gamma \bar n\, \Big(\hat{a}^\dagger\hat{\rho}(m,t) \hat{a}-\frac{1}{2}\{\hat{a} \hat{a}^\dagger,\hat{\rho}(m,t)\}\Big) ,
\end{align}
since the gain term $\mathcal J_e \hat\rho = \gamma (\bar n + 1)\, \hat a \hat\rho \hat a^\dagger$ in the Lindblad equation [Eq.~\eqref{eq:lindblad}] is responsible for photon emissions.

Following the framework of full counting statistics \cite{LevitovJMathPhys1996,BagretsPhysRevB2003,FlindtEPL2004}, we introduce a counting field $s$ by performing a Laplace transformation
\begin{equation}
	\hat \rho(s,t) \equiv \sum_{m=0}^\infty \hat \rho(m,t)\, e^{ms} ,
\end{equation}
which finally transforms the Lindblad equation to
\begin{align} \label{eq:lindblad:cf}
\nonumber
	\frac{d \hat{\rho}(s,t)}{d t} &= -\frac{i}{\hbar}\, [ \hat{H},\hat{\rho}(s,t)] \\
\nonumber
&+ \gamma (\bar n+1)\, \Big(e^s\, \hat{a}\hat{\rho}(s,t) \hat{a}^\dagger-\frac{1}{2}\{\hat{a}^\dagger \hat{a},\hat{\rho}(s,t)\}\Big) \\
&+ \gamma \bar n\, \Big(\hat{a}^\dagger\hat{\rho}(s,t) \hat{a}-\frac{1}{2}\{\hat{a} \hat{a}^\dagger,\hat{\rho}(s,t)\}\Big) .
\end{align}
Since we start counting the emitted photons at time $t=0$, the initial probabilities are $P(m,0) = \delta_{m,0}$, which translates to the initial condition $\hat\rho(s,0) = \hat\rho_0$ for the density matrix.

\subsubsection*{The generating function \texorpdfstring{$\mathcal G(s,q,t)$}{G(s,q,t)}}

Taking the matrix elements $\langle n_1 | \cdots | n_2 \rangle$ of Eq. \eqref{eq:lindblad:cf}, we obtain a set of dynamical equations for the populations $\langle n | \hat \rho(s,t) | n \rangle$ and the coherences $\langle n_1 | \hat \rho(s,t) | n_2 \rangle$ ($n_1 \neq n_2$) of the density matrix.
In order to determine the emission probabilities $P(m,t)$, it is sufficient to determine the dynamics of the populations only, which is possible because the dynamical equations only couple populations to other populations, but not to coherences.
The population dynamics can be fully solved by performing another Laplace transformation
\begin{equation}
	\mathcal G(s,q,t) \equiv \sum_{n=0}^\infty \langle n | \hat\rho(s,t) | n \rangle\, e^{nq} ,
\end{equation}
where $q$ is the variable conjugate to the populations, and thus recasting the system of dynamical equations into a single partial differential equation
\begin{equation} \label{eq:pde}
	\partial_t \mathcal{G}(s,q,t) = [f(s,q)+g(q)]\, \partial_q \mathcal{G}(s,q,t) + g(q)\, \mathcal{G}(s,q,t),
\end{equation}
which is Eq.~\eqref{Partial differential equation} in the main text. For the sake of brevity, we have defined the functions
\begin{equation}
	f(s,q) \equiv \gamma(\bar{n}+1)(e^{s-q}-1), \hspace{3mm}g(q) \equiv \gamma \bar{n}(e^q-1) .
\end{equation}

\subsubsection*{Thermal equilibrium}

At long times after bringing the system in contact with the reservoir, the system will assume the thermal equilibrium state $\hat \rho_{\textrm{eq}} \equiv e^{-\beta \hat H} / Z$, where $Z \equiv \tr e^{-\beta \hat H}$. The probability $P_{\textrm{eq}}(n) \equiv \langle n | \hat\rho_{\textrm{eq}} | n \rangle$ for the cavity to be populated with $n$ photons is then given by the Boltzmann distribution
\begin{align} \label{eq:equilibrium}
\nonumber
&	P_{\textrm{eq}}(n) = \frac 1 Z\, e^{-\beta\hbar\omega_0\, (n + 1/2)}, \\
&Z = \sum_{n=0}^\infty e^{-\beta\hbar\omega_0\, (n + 1/2)} = \frac{1}{e^{\beta\hbar\omega_0/2} - e^{-\beta\hbar\omega_0/2}} .
\end{align}
For later convenience, we also calculate the Laplace transform
\begin{equation} \label{eq:equilibrium:G}
	\mathcal G_{\textrm{eq}}(q) \equiv \sum_{n=0}^\infty P_{\textrm{eq}}(n)\, e^{nq}
		= \frac{1}{1 + \bar n(1 - e^q)} .
\end{equation}

\vspace{2em}
\section{Derivation of the moment generating function [Eq.~\texorpdfstring{\eqref{MGF}}{(\ref*{MGF})}]} \label{sec:mgf}

The general solution of Eq.~\eqref{eq:pde} has the form
\begin{equation} \label{eq:pde:solution}
	\mathcal G(s,q,t) = \mathcal G_0(s,q,t)\, \mathcal G_\ast(s,q,t) ,
\end{equation}
where $\mathcal G_0(s,q,t)$ is the general solution of the homogeneous equation
\begin{equation} \label{eq:pde:homogeneous}
	\partial_t \mathcal{G}(s,q,t) = [f(s,q)+g(q)]\, \partial_q \mathcal{G}(s,q,t)
\end{equation}
and $\mathcal G_\ast(s,q,t)$ is a particular solution of Eq.~\eqref{eq:pde}.

\subsubsection*{Homogeneous solution}

The homogeneous solution is obtained using the method of characteristics \cite{RenardyRogers}. We find the curves $q_Q(s,t)$ along which the solutions $\mathcal G_0(s,q,t)$ are constant,
\begin{equation} \label{eq:method_of_characteristics}
	\frac{d}{dt} \mathcal G_0[s,q_Q(s,t),t] = 0,
\end{equation}
where $Q = q_Q(s,0)$ specifies the initial condition of each curve.
The general homogeneous solution is then
\begin{equation} \label{eq:pde:hom_solution}
	\mathcal G_0(s,q,t) = F(s,Q(s,q,t)),
\end{equation}
where $Q(s,q,t)$ is the inverse to the equation $q=q_Q(s,t)$ and $F(s,q)$ specifies the initial data.

From the condition \eqref{eq:method_of_characteristics} of characteristics, we obtain the ordinary differential equation
\begin{align}
\nonumber
	\frac{dq_Q}{dt} &= -f(s,q_Q)-g(q_Q) \\
&= \gamma \left( 2\bar{n} + 1 - \bar n\, e^{q_Q} - (\bar n + 1)e^s\,e^{-q_Q} \right)
\end{align}
for $q_Q(s,t)$, which can be solved by separation of variables:
\begin{align} \label{eq:pde:t}
\nonumber
	\gamma t
		& = \int \limits_Q^{q_Q} \frac{dx}{ 2\bar{n} + 1 - \bar n\, e^x - (\bar n + 1)e^s\,e^{-x} }\\
		&= \frac{1}{\xi} \ln\frac{\xi - 1 - 2\bar{n}(1-e^x)}{\xi + 1 + 2\bar{n}(1-e^x)}\Bigg|_{x=Q}^{x=q_Q} ,
\end{align}
where $\xi = \sqrt{1 - 4\bar n (1 + \bar n) (e^s - 1)}$.
Solving this result for $Q$, we arrive at the result
\begin{align} \label{eq:pde:Q}
\nonumber
	&Q(s,q,t) = \ln\bigg \{
		\frac{2\bar{n}+1}{2\bar n} +\\
		&\frac{\xi}{2\bar n}\,
		\frac{
			(\xi-1-2\bar{n}[1-e^q])-e^{\xi \gamma t}(\xi+1+2\bar{n}[1-e^q])
		}{
			(\xi-1-2\bar{n}[1-e^q])+e^{\xi \gamma t}(\xi+1+2\bar{n}[1-e^q])
		} \bigg \} ,
\end{align}
which, together with Eq.~\eqref{eq:pde:hom_solution}, solves the homogeneous equation.

\subsubsection*{Particular solution}

To find one particular solution, we make the ansatz of a time-independent solution $\mathcal G_\ast(s,q,t) = \mathcal G_\ast(s,q)$.
Plugging the ansatz into Eq.~\eqref{eq:pde} yields
\begin{widetext}
\begin{align}
\nonumber
	\ln \mathcal G_\ast(s,q)
		&= \int \limits_0^q \frac{-g(x)\, dx}{f(s,x)+g(x)} = -\frac \gamma 2 \int\limits_0^q \frac{dx}{f(s,x)+g(x)} - \frac 1 2 \int\limits_0^q 1\, dx
- \frac 1 2 \int\limits_0^q \frac{d(f(s,x)+g(x))}{dx} \frac{dx}{f(s,x)+g(x)} \\
		&= \frac{\gamma t}{2} \Big|_{Q=0} - \frac 1 2 \ln \frac{e^q\, [f(s,q)+g(q)]}{f(s,0)+g(0)} , \label{eq:pde:part_solution}
\end{align}
\end{widetext}
where $\gamma t$ is short for the expression in Eq.~\eqref{eq:pde:t}.

\subsubsection*{General solution and initial value}

The last remaining step is to express the function $F$ appearing in the homogeneous solution \eqref{eq:pde:hom_solution} in terms of the initial condition $\hat \rho_0$ or, equivalently, the generating function $\mathcal G(s,q,0)$ (which does not depend on $s$).
Evaluating Eq.~\eqref{eq:pde:solution} at time $t=0$, we find that $F$ is given as $F(s,q) = \mathcal G(s,q,0) / \mathcal G_\ast(s,q)$.
Therefore, we can write
\begin{equation}
	\mathcal G(s,q,t) = \mathcal G(s,Q,0)\, \frac{\mathcal G_\ast(s,q)}{\mathcal G_\ast(s,Q)} ,
\end{equation}
where $Q$ is short for the function $Q(s,q,t)$ given in Eq.~\eqref{eq:pde:Q}.
Plugging in Eq.~\eqref{eq:pde:part_solution}, we obtain the full solution
\begin{equation} \label{eq:pde:final_solution}
	\mathcal G(s,q,t) = \mathcal G(s,Q,0)\, e^{\gamma t/2} \sqrt{\frac{e^Q \left[ f(s,Q)+g(Q) \right]}{e^q \left[ f(s,q)+g(q) \right]}} .
\end{equation}

\subsubsection*{Moment generating function}

The moment generating function $\mathcal M(s,t) \equiv \tr \hat\rho(s,t) = \mathcal G(s,0,t)$ is directly obtained from Eq.~\eqref{eq:pde:final_solution}.
Plugging in $q=0$ gives (here $Q_0 \equiv Q(s,0,t)$)
\begin{align}
	Q_0 &= \ln\left\{
			\frac{2\bar n + 1}{2\bar n}
			-\frac{\xi}{2\bar n} \frac{\xi \sinh\!\big[ \frac{\xi\gamma t}{2} \big] + \cosh\!\big[ \frac{\xi\gamma t}{2} \big]}{\xi \cosh\!\big[ \frac{\xi\gamma t}{2} \big] + \sinh\!\big[ \frac{\xi\gamma t}{2} \big]}\right\}
\end{align}
and
\begin{align}
	\mathcal M(s,t) &= \mathcal G(s, Q_0, 0)\, \frac{\xi\, e^{\gamma t / 2}}{\xi \cosh\!\big[ \frac{\xi\gamma t}{2} \big] + \sinh\!\big[ \frac{\xi\gamma t}{2} \big]} , \label{eq:mgf}
\end{align}
where $\xi = \sqrt{1 - 4\bar n (1 + \bar n) (e^s - 1)}$ as above.

If the system is initially thermalized, we can plug in $\mathcal G(s, q, 0) = \mathcal G_{\textrm{eq}}(q)$ as given in Eq.~\eqref{eq:equilibrium:G} and readily obtain
\begin{equation} \label{eq:mgf:final}
	\mathcal M(s,t) = \frac{2\xi\, e^{\gamma t / 2}}{2\xi \cosh\!\big[ \frac{\xi\gamma t}{2}\big] + (1+\xi^2) \sinh\!\big[ \frac{\xi\gamma t}{2}\big]}.
\end{equation}
This equation is identical to Eq.~\eqref{MGF} in the main text.

From the moment generating function, we can for example determine the average emission current $\langle J_e \rangle = \partial_s \mathcal M(s,t) \big|_{s=0} / t$.
Using $\xi = 1 - 2\bar n(1 + \bar n) s + \mathcal O(s^2)$, we expand the moment generating function in powers of $s$, $\mathcal M(s,t) = 1 + \gamma \bar n(1 + \bar n) s t + \mathcal O(s^2)$ and read off
\begin{equation}
	\langle J_e \rangle = \gamma \bar n (1 + \bar n) .
\end{equation}

\subsubsection*{Long time limit}

For long times $\gamma t \gg 1$, we can approximate $\cosh\!\big[ \frac{\xi\gamma t}{2}\big] \simeq \sinh\!\big[ \frac{\xi\gamma t}{2}\big] \simeq e^{\xi\gamma t / 2}$.
Applying this to Eq.~\eqref{eq:mgf:final} yields the moment generating function in the long time limit,
\begin{equation}
	\mathcal M(s,t) \simeq \frac{4\xi}{(1+\xi)^2}\, e^{-\gamma t (\xi - 1) / 2} ,
\end{equation}
as well as the cumulant generating function
\begin{align}
\nonumber
	\Theta(s) &\equiv \lim_{t \to \infty} \frac{\ln \mathcal M(s,t)}{t} = -\frac{\gamma}{2} (\xi - 1) \\
&= \frac \gamma 2 \left( 1 - \sqrt{1 - 4 (e^s - 1) \bar n (1 + \bar n)} \right).
\end{align}

\section{Understanding the waiting time distribution [Eq.~\texorpdfstring{\eqref{WTD}}{(\ref*{WTD})}] with Bayes' theorem}\label{Appendix C}

We use the moment generating function in Eq.~\eqref{eq:mgf:final} to calculate the waiting time distribution
\begin{equation}
	\mathcal W(\tau) = \langle \tau \rangle\, \partial_\tau^2 \Pi(\tau) ,
\end{equation}
where $\langle \tau \rangle = 1 / \langle J_e \rangle$ is the mean waiting time and $\Pi(\tau) =\mathcal M(-\infty,t) =P(n=0,t)$ is the so-called idle-time probability.
With $\langle J_e \rangle = \gamma \bar n (1 + \bar n) \equiv \Gamma$, we obtain
\begin{equation} \label{eq:wtd}
	\mathcal W(\tau) = \! \Gamma\gamma\bar\gamma \frac{\gamma +\! 6\Gamma + \! (\gamma+\! 2\Gamma) \cosh[\bar\gamma \tau] +\! \bar\gamma \sinh[\bar\gamma \tau]}{\left( \bar\gamma \cosh\!\big[ \frac{\bar\gamma \tau}{2} \big] + (\gamma + 2\Gamma) \sinh\!\big[ \frac{\bar\gamma \tau}{2} \big] \right)^3} e^{\frac{\gamma \tau}{2}},
\end{equation}
for the waiting time distribution (WTD), where we have defined $\bar\gamma = \gamma (1 + 2\bar n)$. This is the same equation as Eq.~\eqref{WTD} in the main text.

Expanding the WTD in $\tau$, we find that it equals
\begin{equation} \label{eq:wtd:short_times}
	\mathcal W(\tau) = 2\Gamma\, e^{-\gamma [ 6\bar n (\bar n + 1) + 1] \tau / 2} + \mathcal O(\tau^2)
\end{equation}
for small waiting times.
The leading order at long times is
\begin{equation} \label{eq:wtd:long_times}
	\mathcal W(\tau) \simeq \frac{4\, \Gamma\gamma\bar\gamma}{(\gamma + \bar\gamma + 2\Gamma)^2}\, e^{-\gamma \bar n \tau} .
\end{equation}

\subsubsection*{Consecutive emissions at \texorpdfstring{$t=0$}{t=0}}

From Eq.~\eqref{eq:wtd:short_times}, we find that the probability for a second photon emission immediately after the first is
\begin{equation}
	{W}(0)\, d\tau = 2\Gamma\, d\tau .
\end{equation}
Note that $\mathcal W(0) = \Gamma g^{(2)}(0)$ as shown in Ref. [\citen{PhysRevA.39.1200}], where $g^{(2)}(t)$ is Glauber's second degree of coherence which we will calculate in App.~\ref{sec:g2}.
The functions $\mathcal W(\tau)$ and $\Gamma g^{(2)}(\tau)$ do not agree at first order in $\tau$, however.

The finite value at $\tau = 0$ stems from the fact that the photon cavity can contain many photons at the same time and thus emit several photons within an arbitrarily short time, without a need for particle absorption in between each event.
The waiting time distribution does therefore not display the suppression at short times that WTDs typically display for the emission statistics from single fermionic modes, such as a quantum dot with a single resonance level.

In fact, the cavity has an emission rate that is enhanced by a factor of $2$ at $\tau = 0$ compared to the average emission rate $\langle J_e \rangle = \Gamma$.
This enhancement can be understood from Bayes' theorem.
The conditional probability $P(n|E_0)$ of having $n$ photons in the cavity directly after a photon emission event $E_0$ at time $t=0$ is given by
\begin{equation} \label{eq:bayes}
	P(n|E_0) = P(E_0|n+1)\, \frac{P_{\textrm{eq}}(n+1)}{\Gamma\, dt} = \frac{n+1}{\bar{n}}\, P_{\textrm{eq}}(n+1) ,
\end{equation}
where $P_{\textrm{eq}}(n)$ is the probability for the cavity to be filled with $n$ photons in thermal equilibrium given in Eq.~\eqref{eq:equilibrium}.
We here used that the probability $P(E_0|n+1)$ for the emission event $E_0$, given that there are $n + 1$ photons in the cavity just before, is
\begin{align}
\nonumber
	P(E_0|n+1) &= \gamma (\bar n + 1) \tr\!\left[ a | n+1 \rangle\!\langle n+1 | a^\dagger \right]\, dt \\
&= \gamma(\bar n + 1)\, (n+1)\, dt .
\end{align}
From Eq.~\eqref{eq:bayes}, it follows that the expected number of photons in the cavity after an emission event is exactly twice as large compared to the steady state,
\begin{equation}
	\sum_{n=0}^\infty n P(n | E_0) = 2\bar n
	.
\end{equation}
This result explains our previous observation of the emission rate enhancement, since the emission rate can be calculated as
\begin{equation}
	\Gamma_c(0) = \sum_{n=0}^\infty \gamma (\bar n + 1)\, n P(n | E_0) = 2 \Gamma ,
\end{equation}
where, as in the main text, $\Gamma_c(\tau)$ is the conditional emission rate at a time $\tau$ after the last emission.

\subsubsection*{Conditional emission at finite times}

The conditional emission rate $\Gamma_c(\tau)$ is generally defined as the conditioned rate for emission events $E_\tau$ at time $t = \tau$, given that there was an emission event $E_0$ at time $t = 0$ and no other emissions in between.
By the Kolmogorov definition of conditional probability, this means that
\begin{equation}
	\Gamma_c(\tau) = \frac{\mathcal W(\tau)}{1 - \int_0^\tau \mathcal W(u)\, du} .
\end{equation}
As shown in Fig. 2 (c) in the main text, the emission rate exhibits an enhancement at short times and a suppression at long times compared to its average value $\Gamma$.

From this emission rate, we can calculate the conditional probabilities $P(n | E_0, \tau)$ of having $n$ photons in the cavity at a time $\tau$ after the last emission event.
The probabilities $P(n | E_0, 0)$ are equal to the previously derived $P(n | E_0)$, see Eq.~\eqref{eq:bayes}.
To find the dynamics of these probabilities, we use again the definition of conditional probability, obtaining
\begin{equation}
	P(n | E_0, \tau + d\tau) = \frac{\Pi_n(d\tau | E_0, \tau)}{\Pi(d\tau | E_0, \tau)} .
\end{equation}
Here $\Pi(d\tau | E_0, \tau) = 1 - \Gamma_c(\tau)\, d\tau$ is the conditioned idle-time probability for no emission events during the time $d\tau$, given the emission event $E_0$ and that there was no other emission between times $t=0$ and $t=\tau$.
The quantity $\Pi_n(d\tau | E_0, \tau)$ is the conditioned idle-time probability for no emission events during $d\tau$ \emph{and} for the cavity to contain $n$ photons at time $t=\tau+d\tau$.
It can be calculated as
\begin{align}
\nonumber
	\Pi_n(d\tau | E_0, \tau) &= P(n | E_0, \tau) \left[ 1 - \gamma_{n \to n+1}\, d\tau - \gamma_{n \to n-1}\, d\tau \right] \\
&+ P(n-1 | E_0, \tau)\, \gamma_{n-1 \to n}\, d\tau ,
\end{align}
since $d\tau$ is an infinitesimally short time and there can be at most only one emission or absorption event during this time.
Here $\gamma_{n \to n'}$ denotes the rate at which the number of cavity photons changes from $n$ to $n'$.
More specifically, $\gamma_{n \to n+1} = \gamma \bar n (1+n)$ and $\gamma_{n \to n-1} = \gamma (\bar n + 1) n$.
Together, we obtain a system
\begin{align} \label{eq:conditioned_probabilities}
\nonumber
	&\partial_\tau P(n | E_0, \tau) = \Gamma_c(\tau) P(n | E_0, \tau) + \gamma \bar n n P(n - 1 | E_0, \tau)\\
&\qquad\qquad\quad - \gamma \left[ \bar n (1+n) + (\bar n + 1) n \right] P(n | E_0, \tau)
\end{align}
of differential equations for the conditional probabilities that can be solved at least numerically, see Fig.~\ref{WTDs}~(d).
As a consistency check, we can calculate $\Gamma_c(\tau)$ as well as the waiting time distribution $\mathcal W(\tau)$ from the probabilities $P(n | E_0, \tau)$ and get back the previous results:
\begin{align}
	\Gamma_c(\tau) &= \sum_{n=0}^\infty \gamma (\bar n + 1)\, n P(n | E_0, \tau) , \\
	\mathcal W(\tau) &= e^{-\int_0^\tau \Gamma_c(u)\, du}\, \Gamma_c(\tau) .
\end{align}

\subsubsection*{Long time behavior}

In the long time limit, we expect the probability distribution to reach a steady state.
Looking for such a solution, we set $\partial_\tau P(n | E_0, \tau)$ to zero in Eq.~\eqref{eq:conditioned_probabilities} and obtain
\begin{equation} \label{eq:conditioned_probabilities:long_time}
	\frac{P_{\textrm{eq}}(n|E_0,\tau)}{P_{\textrm{eq}}(n-1|E_0,\tau)} = \frac{\bar n}{1 + 2\bar n}
\end{equation}
for the steady state.
We used that the waiting time distribution for long times [see Eq.~\eqref{eq:wtd:long_times}] resembles a Poissonian process with emission rate $\gamma \bar n$,
\begin{equation}
	\lim_{\tau \to \infty} \Gamma_c(\tau) = \gamma \bar n .
\end{equation}

Noting that for a Boltzmann distribution $\frac{P_{\textrm{eq}}(n)}{P_{\textrm{eq}}(n-1)} = \frac{\bar n}{1 + \bar n}$, we find that the conditional probability distribution for long times is a Boltzmann distribution as well, albeit with modified $\tilde{n} = \frac{\bar n}{1 + \bar n}$.
In other words, $\tilde n$ is the expected number of photons in the cavity a long time after the last emission event.
As one would expect, $\tilde{n}$ is always between $0$ (for $\bar n = 0$) and $1$ (for $\bar n \to \infty$):
	If there has not been any emission for a very long time, we expect the average number of photons in the cavity to be less than one.
Note also that $\tilde{n} \approx \bar{n}$ for low temperatures.

\section{Calculation of the \texorpdfstring{$\bm{g^{(2)}}$}{g2}-function [Eq.~\texorpdfstring{\eqref{SDC}}{(\ref*{SDC})}]} \label{sec:g2}

From the moment generating function in Eq.~\eqref{eq:mgf:final} we calculate the noise spectrum $S(\omega)$ of the emission current using MacDonald's formula \cite{0034-4885-12-1-304},
\begin{equation}
S(\omega) = \omega \int \limits_0^\infty dt \sin(\omega t)\frac{d}{dt}\langle\!\langle m^2\rangle\!\rangle(t) =  \Gamma\left(1+2\frac{\Gamma\gamma}{\gamma^2+\omega^2}\right),
\label{Noise spectrum}
\end{equation}
where
\begin{align}
\nonumber
	\langle\!\langle m^2\rangle\!\rangle(t)& = \frac{d^2}{ds^2}\mathcal{M}(s,t)\Big|_{s=0} =\bar{n}(1+\bar{n}) \Big[ 2\bar{n}(1+\bar{n})e^{-\gamma t} \\
&+ \gamma t + \bar{n}(1+\bar{n}) \big( \gamma t [2+\gamma t] - 2 \big) \Big] .
\end{align}
The first part in Eq.~\eqref{Noise spectrum} is due to self-correlations. From the definitions of $S(\omega)$ and $g^{(2)}(\tau)$, it follows that \cite{PhysRevB.85.165417} $g^{(2)}(\tau)$ is related to the inverse Fourier transform of the noise spectrum without the self-correlation part as
\begin{equation} \label{eq:g2}
	g^{(2)}(\tau) = 1+\frac{1}{2\pi \Gamma^2}\int \limits_{-\infty}^\infty d\omega\, e^{-i\omega \tau} \left[S(\omega)-\Gamma \right] = 1+e^{-\gamma |\tau|},
\end{equation}
which provides the result given in Eq.~\eqref{SDC} in the main text. An important observation is that the $g^{(2)}$-function does not depend on the bath temperature.
Therefore, it cannot be possible in general to derive the waiting time distribution from the $g^{(2)}$-function alone.

\subsubsection*{Photon emission is not a renewal process}

For renewal processes, for which subsequent waiting times are uncorrelated, the WTD can be derived from the $g^{(2)}$-function and $\Gamma = \langle J_e \rangle$ using the relation
\begin{equation} \label{eq:renewal}
	\Gamma\, g^{(2)}(s) = \mathcal{W}(s)/[1-\mathcal{W}(s)]
\end{equation}
for the Laplace transformed $g^{(2)}(s)$ and $\mathcal{W}(s)$ \cite{PhysRevA.39.1200}.

Here, we compute the WTD $\mathcal{W}_\textrm{re}(\tau)$ that one obtains from applying this formula to the present case, therefore assuming that the photon emission process is a renewal process.
We show that this distribution is different from the correct waiting time distribution given in Eq.~\eqref{eq:wtd}, thus showing explicitly that the process is \emph{not} a renewal process.

The Laplace transform of the $g^{(2)}$-function given in Eq.~\eqref{eq:g2} is $g^{(2)}(s) = \frac{1}{s}+\frac{1}{s+\gamma}$, using Eq.~\eqref{eq:renewal} we get
\begin{equation}
	\mathcal{W}_\textrm{re}(s) = \frac{\Gamma\, g^{(2)}(s)}{1 + \Gamma\, g^{(2)}(s)} = \frac{\Gamma(2s+\gamma)}{s(s+\gamma)+\Gamma(2s+\gamma)}.
\end{equation}
Performing an inverse Laplace transform, we obtain
\begin{equation} \label{WTD under renewal condition}
	\mathcal{W}_\textrm{re}(\tau) = \frac{\Gamma}{\tilde{\Gamma}}
			\left[ \tilde{\Gamma} (e^{\tilde{\Gamma}\tau}+1) - 2\Gamma (e^{\tilde{\Gamma}\tau}-1) \right]
		e^{-\frac{1}{2}(\gamma+2\Gamma+\tilde{\Gamma})\tau},
\end{equation}
with $\tilde{\Gamma} = \sqrt{\gamma^2+4\Gamma^2}=\gamma\sqrt{1+4n^2(1+n)^2}$.
This WTD is evidently different from the one given in Eq.~\eqref{eq:wtd}, showing that the emission statistics is a non-renewal process.
At short waiting times $\tau$, the expression~\eqref{WTD under renewal condition} reduces to
\begin{equation}
	\mathcal{W}_\textrm{re}(\tau) \simeq 2\Gamma\, e^{-\gamma [ 4\bar n (\bar n + 1) + 1] \tau / 2} ,
\end{equation}
which can be compared to the short-time WTD given in Eq.~\eqref{eq:wtd:short_times}, $\mathcal{W}(\tau) \simeq 2\Gamma\, e^{-\gamma [ 6\bar n (\bar n + 1) + 1] \tau / 2}$.
The WTD of the cavity decays faster as a result of the bunching effect.
We see that for low temperatures, for which $\bar n \to 0$, the two distributions give the same result.
In that case, it is very unlikely that there is more than one photon in the cavity and thus the cavity returns to the same state after every emission; this is a renewal process.

\section{Large-deviation statistics of the emission current}
\label{Appendix E}

Here we discuss the long-time statistics of the emission current, described by the cumulant generating function $\Theta(s) \equiv \lim_{t \to \infty} \frac{\ln \mathcal M(s,t)}{t}$.
As shown in App.~\ref{sec:mgf}, it has the form
\begin{equation} \label{eq:cgf}
	\Theta(s) = \frac \gamma 2 \left( 1 - \sqrt{1 - 4 (e^s - 1) \bar n (1 + \bar n)} \right)
\end{equation}
for the emission current. This is the same equation as Eq.~\eqref{eq:CGFemis} in the main text.

We recall that the moment-generating function $\mathcal M(s,t) \sim e^{\Theta(s) t}$ is defined as
\begin{equation}
	\mathcal M(s,t) = \sum_m P(m,t)\, e^{ms}
\end{equation}
where $P(m,t)$ is the probability to have emitted $m$ photons at time $t$.
This relation allows us to extract the probability $P(J_e,t)$ for having an average emission current $J_e = m/t$ during a measurement time $t$ as a Fourier coefficient of the moment generating function, it is
\begin{align}
\nonumber
	P(J_e,t)& = \frac{1}{2\pi i} \int\limits_{-i\pi}^{i\pi} ds\, \mathcal M(s,t)\, e^{-ms} \\
&= \frac{1}{2\pi i} \int\limits_{-i\pi}^{i\pi} ds\, e^{t [\Theta(s) - sJ_e]} .
\end{align}
In the long-time limit, this integral can be solved using the saddle-point approximation.
Let $s_0$ be the solution to the saddle-point equation
\begin{equation} \label{eq:saddle_point}
	\Theta'(s_0) = J_e ,
\end{equation}
then the exponent of the integral equals $t \big[ \Theta(s_0) - s_0 J_e + \frac 1 2 \Theta^{\prime\prime}(s_0) (s - s_0)^2 \big]$ to second order.
The integral can be performed explicitly, and after taking the limit of large $t$ we are only left with
\begin{equation} \label{eq:ldf}
	\frac{\ln[P(J_e,t)]}{t} \simeq \Theta(s_0) - s_0 J_e
\end{equation}
up to terms of order $\ln[t]/t$, note that $s_0$ does not depend on $t$. This is Eq.~\eqref{Saddle-point approximation} in the main text.
The quantity $\lim_{t \to \infty} \ln[P(J_e,t)] / t$ is called the large deviation function.

Solving Eq.~\eqref{eq:saddle_point}, we obtain
\begin{equation} \label{eq:saddle_point:solution}
	s_0 = \ln\left[ \frac{J_e}{\gamma\Gamma} \left( \sqrt{ 4J_e^2 + \gamma^2 + 4\gamma\Gamma } - 2J_e \right) \right] ,
\end{equation}
and plugging this back into Eq.~\eqref{eq:ldf} gives the final result
\begin{align} \label{eq:ldf:full}
\nonumber
	&\frac{\ln[P(J_e,t)]}{t} = \frac{\gamma}{2} + J_e - \frac 12 \sqrt{4J_e^2 + \gamma^2 + 4\gamma\Gamma} \\
&+ J_e \ln\left[ \frac{\gamma\Gamma}{J_e (\sqrt{4J_e^2 + \gamma^2 + 4\gamma\Gamma} - 2J_e)} \right] .
\end{align}
For examples illustrating this distribution, see Fig.~\ref{Long-time statistics}~(a).

\subsubsection*{Large \texorpdfstring{$J_e$}{Je} limit}

For $J_e \gg \gamma,\Gamma$, we obtain from Eq.~\eqref{eq:saddle_point:solution} that
\begin{equation}
	s_0 \simeq s_c \equiv \ln\left[ 1+\frac{\gamma}{4\Gamma} \right] ,
\end{equation}
where $s_c$ is the locus of the square-root singularity of the cumulant-generating function $\Theta(s)$ with $\Theta(s_c) = \gamma/2$.
Plugging back into Eq.~\eqref{eq:ldf}, we obtain
\begin{equation}
	\frac{\ln[P(J_e,t)]}{t} \simeq \frac \gamma 2 - \ln\left[ 1+\frac{\gamma}{4\Gamma} \right] J_e .
\end{equation}
Thus the tail of the probability distribution decays exponentially.

In the limit of high or low temperatures, this expression can be simplified further.
For high temperatures $\bar n \gtrsim 1$, we approximate the slope as $\ln\left[ 1+\frac{\gamma}{4\Gamma} \right] \approx (\beta\hbar\omega_0)^2 / 4$, resulting in
\begin{equation}
	\frac{\ln[P(J_e,t)]}{t} \simeq \frac{\gamma}{2} - \frac{(\beta \hbar \omega_0)^2}{4} J_e ,
\end{equation}
whereas for low temperatures $\bar n \ll 1$, we use that $\ln\left[ 1+\frac{\gamma}{4\Gamma} \right] \approx \beta\hbar\omega_0 - \ln 4$.
At even lower temperatures, we can neglect also the constant offset, obtaining
\begin{equation}
	\frac{\ln[P(J_e,t)]}{t} \simeq \frac{\gamma}{2} - \beta \hbar \omega_0\, J_e,
\end{equation}
which is Eq.~\eqref{eq:LDFapprox} in the main text.

\subsubsection*{Poissonian limit}

In the other limit, $J_e\ll \gamma$, we obtain from Eq.~\eqref{eq:ldf:full}
\begin{equation}
	\frac{\ln[P(J_e,t)]}{t} \simeq \left( J_e - \gamma \bar n \right) - J_e \ln\left[J_e \frac{\bar\gamma}{\gamma\Gamma} \right].
\end{equation}
For $\bar{n}\ll 1$, this is a Poissonian distribution corresponding to the CGF
\begin{equation}
	\Theta_{\text{poiss}}(s) = \Gamma \left( e^s - 1 \right) ,
\end{equation}
which is exactly what Eq.~\eqref{eq:cgf} reduces to in this limit.

\section{Generalization to multiple heat baths} \label{sec:generalization}

Here we generalize the previous results to multiple heat baths. We consider a cavity coupled to $N$ heat baths, each with a coupling constant $\gamma_i$ and an inverse temperature $\beta_i$. The Lindblad equation is
\begin{align} \label{eq:multiplelindblad}
\nonumber
	\frac{d \hat{\rho}}{d t} &= -\frac{i}{\hbar}\, [ \hat{H},\hat{\rho}] + \sum_{i=1}^N \bigg[\gamma_i (\bar n_i+1)\, \Big( \hat{a}\hat{\rho} \hat{a}^\dagger-\frac{1}{2}\{\hat{a}^\dagger \hat{a},\hat{\rho}\}\Big) \\
&+ \gamma_i \bar n_i\, \Big(\hat{a}^\dagger\hat{\rho} \hat{a}-\frac{1}{2}\{\hat{a} \hat{a}^\dagger,\hat{\rho}\}\Big) \bigg] ,
\end{align}
which is a direct generalization of Eq.~\eqref{eq:lindblad}.
As before, $\bar n_i = \frac{1}{e^{\beta_i \hbar\omega_0} - 1}$ is the Bose-Einstein factor corresponding to the mode $\omega_0$ of the $i$-th reservoir.

To keep track of the number $m^+_i$ ($m^-_i$) of photons emitted into (absorbed from) heat bath $i$, we introduce the $m^\pm_i$-resolved density matrices $\hat{\rho}(m^{\pm}_i,t)$, so that $P(m^{\pm}_i,t) = \tr \hat{\rho}(m^{\pm}_i,t)$ is the probability of having emitted/absorbed $m^\pm_i$ photons to/from heat bath $i$. Analogously to the single-bath case, we then perform a Laplace transformation
\begin{equation}
	\hat{\rho}(s^\pm_i,t) = \sum_{\mathclap{m^\pm_i=0}}^\infty\, \hat{\rho}(m^{\pm}_i,t)\, e^{\sum_{i=1}^N(m^+_is^+_i-m^-_is^-_i)}
\end{equation}
with two counting fields per bath, $s^+_i$ for absorption and $s^-_i$ for emission.
We then introduce the generating function $\mathcal{G}(s^\pm_i,q,t) = \sum_{n=0}^\infty \langle n| \hat{\rho}(s^\pm_i,t) |n\rangle\, e^{nq}$ and obtain the partial differential equation
\begin{align}
\nonumber
	\partial_t \mathcal{G}(s^\pm_i,q,t) &= [f(s^+_i,q)+g(s^-_i,q)]\, \partial_q \mathcal{G}(s^\pm_i,q,t) \\
&+ g(s^-_i,q)\, \mathcal{G}(s^\pm_i,q,t) ,
\end{align}
with
\begin{align}
	f(s^+_i,q) &\equiv \sum_{i=1}^N \gamma_i (\bar{n}_i+1)\left(e^{s^+_i-q}-1\right)\\
	\nonumber \text{and} \\
	g(s^-_i,q) &\equiv \sum_{i=1}^N \gamma_i \bar{n}_i\left(e^{q-s^-_i}-1\right).
\end{align}
From the method of characteristics, we get the solution
\begin{align} \label{eq:generating_function:general}
\nonumber
\mathcal G(s^\pm_i,q,t) &= \mathcal G(s^\pm_i,Q,0)\, e^{\gamma_\Sigma t/2} \\
&\times \sqrt{\frac{e^Q \left[ f(s^+_i,Q)+g(s^-_i,Q) \right]}{e^q \left[ f(s^+_i,q)+g(s^-_i,q) \right]}},
\end{align}
with $\gamma_\Sigma = \sum_{i=1}^N \gamma_i$ and
\begin{widetext}
\begin{equation}
	Q(s,q,t) = \ln\left \{
		\frac{2\bar{n}+1}{2n^-} +
		\frac{\xi}{2n^-}\,
		\frac{
			(\xi-[1+2\bar{n}]+2{n}^-e^q)-e^{\xi \gamma_\Sigma t}(\xi+[1+2\bar{n}]-2{n}^-e^q])
		}{
			(\xi-[1+2\bar{n}]+2{n}^-e^q)+e^{\xi \gamma_\Sigma t}(\xi+[1+2\bar{n}]-2{n}^-e^q])
		} \right \} .
\end{equation}
\end{widetext}
Similar to before, $\xi$ is defined as $\xi = \sqrt{(2\bar{n}+1)^2-4n^-(n^++1)}$, where $\bar{n} = \sum_{i=1}^N \frac{\gamma_i}{\gamma_\Sigma}\bar{n}_i$ is the average number of photons in the cavity in the steady state, and $n^-=\sum_{i=1}^N\frac{\gamma_i}{\gamma_\Sigma}\bar{n}_ie^{-s^-_i}$ and $n^++1=\sum_{i=1}^N\frac{\gamma_i}{\gamma_\Sigma}(\bar{n}_i+1)e^{s^+_i}$.
We focus on the case where the initial state is the steady state, i.e., $\mathcal G(s^\pm_i,q,0) = \mathcal{G}_\textrm{eq}(q)$, see Eq.~\eqref{eq:equilibrium:G}.

\subsubsection*{Emission current statistics}

To generalize the previous results for the emission statistics, we compute the moment generating function (MGF) for photon emission from the cavity to heat bath $i=1$. To this end, we set all counting fields to zero except $s\equiv s^+_1$, the counting field corresponding to emission into heat bath $i=1$.
The moment generating function $\mathcal M(s,t) =\tr \hat{\rho}(s,t) = \mathcal G(s,0,t)$ can be calculated from \eqref{eq:generating_function:general}, it is
\begin{equation} \label{eq:mgf:emission}
	\mathcal M(s,t) = \frac{2\xi\, e^{\gamma_\Sigma t / 2}}{2\xi \cosh\!\big[ \frac{\xi\gamma_\Sigma t}{2}\big] + (1+\xi^2) \sinh\!\big[ \frac{\xi\gamma_\Sigma t}{2}\big]} ,
\end{equation}
where $\xi = \sqrt{1-4\frac{\gamma_1}{\gamma_\Sigma}\bar{n}(1+\bar{n}_1)(e^s-1)}$.
We see that the MGF resembles the one of a single heat bath given in Eq.~\eqref{eq:mgf:final}.

From the MGF we obtain the waiting time distribution
\begin{widetext}
\begin{equation} \label{eq:wtd:emission}
	\mathcal W(\tau) = \Gamma\gamma_\Sigma\bar\gamma \frac{\gamma_\Sigma + 6\Gamma + (\gamma_\Sigma+2\Gamma) \cosh[\bar\gamma \tau] + \bar\gamma \sinh[\bar\gamma \tau]}{\left( \bar\gamma \cosh\!\big[ \frac{\bar\gamma \tau}{2} \big] + (\gamma_\Sigma + 2\Gamma) \sinh\!\big[ \frac{\bar\gamma \tau}{2} \big] \right)^3} e^{\frac{\gamma_\Sigma \tau}{2}},
\end{equation}
\end{widetext}
with $\bar{\gamma} = \gamma_\Sigma \sqrt{1+4\frac{\gamma_1}{\gamma_\Sigma}\bar{n}(1+\bar{n}_1)}$.
Here, $\Gamma = \gamma_1 \bar n (1 + n_1)$ is the average emission rate into the first reservoir and the mean waiting time is $\langle \tau \rangle = \Gamma^{-1}$.

Similarly, we get the $g^{(2)}$-function
\begin{equation}
g^{(2)}(\tau) = 1+e^{-\gamma_\Sigma |\tau|},
\end{equation}
which, again, is temperature independent in contrast to the WTD.

\subsubsection*{Net current statistics}

We consider the net current statistics between the cavity and a heat bath with average occupation number $\bar{n}_c$ and coupling strength $\gamma_c$. The cavity is assumed to be coupled to another heat bath with occupation number $\bar{n}_h$ and coupling strength $\gamma_h$. The MGF of the net current is obtained as $\mathcal M(s,t) =\tr \hat{\rho}(s,t) = \mathcal G(s,0,t)$ (where $s = s^+_c = s^-_c$ and $s^+_h = s^-_h =0$), yielding
\begin{equation}\label{eq:MGFnetcurrent}
	\mathcal M(s,t) = \frac{2\xi\, e^{\gamma_\Sigma t / 2}}{2\xi \cosh\!\big[ \frac{\xi\gamma_\Sigma t}{2}\big] + (1 + \chi^2) \sinh\!\big[ \frac{\xi\gamma_\Sigma t}{2}\big]},
\end{equation}
where
\begin{align}
\nonumber
	\xi &\!=\! \sqrt{1\!-\!4\frac{\gamma_c\gamma_h}{\gamma_\Sigma^2} \big[ (e^s-\!1)(1\!+\!\bar{n}_c)\bar{n}_h+\!(e^{-s}-\! 1)\bar{n}_c(1\!+\!\bar{n}_h) \big]} , \\
	\chi &\!=\! \sqrt{1\!-\!4\frac{\gamma_c\gamma_\Sigma}{\gamma_\Sigma^2} \big[ (e^s-\!1)(1\!+\!\bar{n}_c) \bar{n} +(e^{-s}-\!1) \bar{n}_c (1\!+\!\bar{n})\big]} .
\end{align}
If the temperature of the cold reservoir is very low, the photon current from the cold reservoir into the system goes to zero and this result reduces to the previously derived emission current statistics.
More precisely, if $\bar n_c$ is set to zero in \eqref{eq:MGFnetcurrent}, we obtain back the moment generating function in Eq.~\eqref{eq:mgf:final} describing the emission current into a single heat bath with the decay rate $\gamma = \gamma_c + \gamma_h$ and an effective temperature given by
\begin{equation}
	\bar n_{\text{eff}} = \frac 1 2 \left( \sqrt{ 1 + 4\bar n_h \frac{ \gamma_c \gamma_h }{ (\gamma_c + \gamma_h)^2 } } - 1 \right) .
\end{equation}
Moreover, for $\gamma_c \gg\gamma_h$, these expressions simplify to $\gamma \simeq \gamma_c$ and $\bar n_{\text{eff}} \simeq (\gamma_h/\gamma_c)\bar n_h$.

In the long-time limit, we find the cumulant generating function $\Theta(s) =\lim_{t\rightarrow \infty}  \frac{\ln \mathcal{M}(s,t)}{t}$ for the net current,
\begin{equation}\label{eq:CGFlongtime}
\Theta(s) = \frac{\gamma_c+\gamma_h}{2}\left(1-\sqrt{1-4\frac{\gamma_c\gamma_h}{(\gamma_c+\gamma_h)^2}\kappa(s)}\right)
\end{equation}
with $\kappa(s)\equiv (e^s-1)(1+\bar{n}_c)\bar{n}_h+(e^{-s}-1)\bar{n}_c(1+\bar{n}_h)$, which is Eq.~\eqref{Long-time CGF net current} in the main text.

\section{Derivation of the fluctuation relation [Eq.~\texorpdfstring{\eqref{FR}}{(\ref*{FR})}]}
\label{Appendix G}
To derive a fluctuation relation for the net current in the long time limit, we note that the CGF in Eq.~\eqref{eq:CGFlongtime} fulfills the symmetry property
\begin{equation}
\Theta(s) = \Theta(-s-\sigma),
\end{equation}
with $\sigma = \hbar \omega_0(\beta_c-\beta_h)$ determining the entropy increase per transferred photon. We then obtain the following result for the probability distribution
\begin{align}
\nonumber
P(J,t) &= \frac{1}{2\pi}\! \! \int \limits_{-\infty}^{\infty} e^{\Theta(s)}e^{-sJt}ds = \frac{1}{2\pi}\! \!\int \limits_{-\infty}^{\infty} e^{\Theta(-s-\sigma)}e^{-sJt}ds\\
& = \frac{1}{2\pi}\int \limits_{-\infty}^{\infty} e^{\Theta(s)}e^{(s+\sigma)Jt}ds = P(-J,t)e^{\sigma t J}.
\end{align}
We have thus derived the fluctuation relation
\begin{equation}
\frac{1}{t}\ln\left[\frac{P(J,t)}{P(-J,t)}\right] = \sigma J.
\end{equation}
This is Eq.~\eqref{FR} in the main text.

\section{Relations between equilibrium noise and response coefficients}\label{Appendix H}
From the symmetry property $\Theta(s) = \Theta(-s-\sigma)$, we now also derive the fluctuation-dissipation theorem. For clarity, we will below let $\Theta(s,\sigma)$ have a second argument indicating the dimensionless temperature difference $\sigma$ of the cavity. The average particle current between two heat baths with different temperatures, $\sigma/(\hbar\omega_0) = \beta_c-\beta_h$, is then given by
\begin{align}
\nonumber
\langle I \rangle  &= \partial_s \Theta(s,\sigma)\big|_{s=0} = \partial_s \Theta(-s-\sigma,\sigma)\big|_{s=0} \\
&= -\Theta^{(1,0)}(-\sigma,\sigma),
\end{align}
where the superscripts refers to the number of derivatives with respect to the first and the second argument, respectively. All quantities are evaluated at $s=0$ after the differentiations. Expanding $\langle I \rangle$ in $\sigma$ to second order, we obtain
\begin{eqnarray}
\nonumber
\langle I \rangle \approx -\Theta^{(1,0)}(0,0)+\left[\Theta^{(2,0)}(0,0)-\Theta^{(1,1)}(0,0) \right]\sigma\\
+\left[-\Theta^{(3,0)}(0,0)+2\Theta^{(2,1)}(0,0)-\Theta^{(1,2)}(0,0)\right] \frac{\sigma^2}{2}.\hspace{4mm}
\end{eqnarray}
Using that all odd cumulants are zero in equilibrium, $\Theta^{(n,0)}(0,0) = 0$ for $n=1,3,5,...$, and identifying each prefactor of $\sigma^n / n!$ with $\frac{\partial^n \langle I\rangle}{\partial \sigma^n}\big|_\textrm{eq} = \Theta^{(1,n)}(0,0)$, we obtain the following relations
\begin{equation}
\Theta^{(1,1)}(0,0) = \frac{1}{2}\Theta^{(2,0)}(0,0), \quad \Theta^{(1,2)}(0,0) = \Theta^{(2,1)}(0,0).
\end{equation}

\subsubsection*{Linear regime}
We consider the linear thermal conductance
\begin{align}
\nonumber
&\frac{1}{\hbar\omega_0}G_Q^{(1)} =\frac{1}{\hbar\omega_0} \frac{\partial \langle J\rangle }{\partial \Delta T}\bigg|_{\Delta T=0} = \frac{\partial \langle I \rangle }{\partial \sigma}\bigg|_{\sigma = 0} \frac{\partial \sigma}{\partial \Delta T}\bigg|_{\Delta T=0} \\
&= \Theta^{(1,1)}(0,0)\frac{\partial \sigma}{\partial \Delta T}\bigg|_{\Delta T= 0} =\frac{1}{2}\Theta^{(2,0)}(0,0)\frac{\partial \sigma}{\partial \Delta T}\bigg|_{\Delta T= 0},
\end{align}
where $\langle J \rangle = \hbar \omega_0 \langle I \rangle$ is the heat current. Using that $\frac{\partial \sigma}{\partial \Delta T}\big|_{\Delta T= 0} = \frac{\hbar \omega_0}{k_B T^2}$, we obtain
\begin{equation}
G_Q^{(1)} = (\hbar\omega_0)^2\frac{1}{2k_B T^2}\Theta^{(2,0)}(0,0),
\end{equation}
or,
\begin{equation}
S_Q^\textrm{(eq)} = 2k_BT^2G_Q^{(1)},
\end{equation}
where we have introduced the equilibrium heat noise $S_Q^\textrm{(eq)} = (\hbar \omega_0)^2\Theta^{(2,0)}(0,0)$. This is the fluctuation-dissipation theorem for heat currents, relating the equilibrium noise to the linear thermal conductance.

\subsubsection*{Weakly non-linear regime}
For the weakly non-linear regime, we get
\begin{widetext}
\begin{eqnarray}
\nonumber
G_Q^{(2)}&=& \frac{\partial^2 \langle J\rangle }{\partial \Delta T^2}\bigg|_{\Delta T=0}  = \hbar\omega_0 \frac{\partial^2 \langle I\rangle}{\partial \Delta T^2}\bigg|_{\Delta T=0} = \hbar \omega_0 \left(F^{(1,2)}(0,0)\left[\frac{\partial \sigma}{\partial \Delta T}\right]^2+F^{(1,1)}(0,0)\frac{\partial^2 \sigma}{\partial \Delta T^2} \right)\bigg|_{\Delta T = 0}\\
&=&  \hbar \omega_0 \left(F^{(2,1)}(0,0)\left[\frac{\hbar \omega_0}{k_BT^2}\right]^2\right)\Bigg|_{\Delta T = 0} = \frac{1}{\hbar \omega_0} \frac{\partial S_Q}{\partial \Delta T}\bigg|_{\textrm{eq}}\frac{\partial \Delta T}{\partial \sigma }\bigg|_{\Delta T =0} \left[\frac{\hbar \omega_0}{k_BT^2}\right]^2 = \frac{\partial S_Q}{\partial \Delta T}\bigg|_{\Delta T=0}\frac{1}{k_BT^2},
\end{eqnarray}
\end{widetext}
where we have used $\frac{\partial^2 \sigma}{\partial \Delta T^2}\big|_{\Delta T=0} =0$. We thus arrive at the relation
\begin{equation}
S^{(1)}_Q\equiv\frac{\partial S_Q}{\partial \Delta T}\bigg|_\textrm{eq} = k_BT^2 G_Q^{(2)}.
\end{equation}

\section{Noise power spectrum and the fluctuation--dissipation theorem at finite frequency} \label{Appendix I}

We now consider a setup with a cavity coupled to two heat baths with the same temperature, i.e., the average photon occupation number is $\bar{n} \equiv \bar{n}_c=\bar{n}_h$. From the moment generating function in Eq.~\eqref{eq:MGFnetcurrent}, we then obtain
\begin{eqnarray}
\nonumber
\frac{\langle\! \langle m_{c,h}^2 \rangle\!\rangle(t)}{2\bar{n}(1+\bar{n})} &=& \frac{\gamma_{c,h}^2[1-e^{-t(\gamma_c+\gamma_h)}]+t\gamma_c\gamma_h(\gamma_c+\gamma_h)}{(\gamma_c+\gamma_h)^2},\\
\frac{\langle\! \langle m_cm_h \rangle\!\rangle(t)}{2\bar{n}(1+\bar{n})}  &=& \frac{\gamma_c\gamma_h[1-e^{-t(\gamma_c+\gamma_h)}-t(\gamma_c+\gamma_h)]}{(\gamma_c+\gamma_h)^2},
\end{eqnarray}
where $m_c$ ($m_h$) denotes the number of particles transferred into heat bath $c$ ($h$) over a time $t$. Using MacDonald's formula [see Eq.~\eqref{Noise spectrum}], we obtain the following expression for the spectral densities $S^{c}_Q(\omega)$, $S^{h}_Q(\omega)$ and $S^{ch}_Q(\omega)$ of the particle currents (to the cold and hot baths and the cross term, respectively)
\begin{align}
\nonumber
S^{c,h}_{Q}(\omega) &=S^{\textrm{(eq)}}_Q\left(1+\frac{\gamma_{c,h}}{\gamma_{h,c}}\frac{\omega^2}{(\gamma_c+\gamma_h)^2+\omega^2}\right),\\
\textrm{Re}\left[S^{ch}_{Q}(\omega)\right] &=S^{\textrm{(eq)}}_Q\left(-1+\frac{\omega^2}{(\gamma_c+\gamma_h)^2+\omega^2}\right),
\end{align}
where $S^{\textrm{(eq)}}_Q = 2 \bar{n}(1+\bar{n})\frac{\gamma_c\gamma_h}{\gamma_c+\gamma_h} = 2k_BT^2G_Q^{(1)}$. These equations are identical to Eqs.~\eqref{Eq12} and \eqref{Eq13} in the main text.

Using the continuity equation, $\dot{U}(t) = -[J_Q^c(t)+J_Q^h(t)]$ for the cavity energy and the outgoing heat currents, we write the energy fluctuations as
\begin{equation}
\omega^2S_U(\omega)=S^c_Q(\omega)+S^h_Q(\omega)+2\textrm{Re}[S^{ch}_Q(\omega)].
\label{Continuity equation transformed}
\end{equation}
From this equation, we get
\begin{equation}
S_U(\omega)= 2(\hbar \omega_0)^2\bar{n}(1+\bar{n})\frac{(\gamma_c+\gamma_h)}{(\gamma_c+\gamma_h)^2+\omega^2}
\label{S_U expression}
\end{equation}

\subsubsection*{Linear response}
We consider a perturbed oscillator, with Hamiltonian $\hat{H}(t) = \hat{H}_0 + \hat{H}_1(t)$, where $\hat{H}_0 =\hbar\omega_0\left(\hat{n}+\frac{1}{2}\right)$ is the unperturbed Hamiltonian and $\hat{H}_1(t) = \hat{H}_0 K(t)$ is a weak perturbation, where $K(t)$ determines the modulation. Below we find the susceptibility that relates the response in the cavity energy $\delta U(t) = \hbar \omega_0 \delta n(t)$ to the modulation $K(t)$.

To this end, we first introduce the mean number of cavity photons $\langle n(t)\rangle = \sum_n n P(n,t)$ as a function of time. From the Lindblad equation we have
\begin{equation}
\frac{d\langle n(t)\rangle}{dt} = \gamma_c\left[n_c(t)-\langle n(t)\rangle\right]+\gamma_h\left[n_h(t)-\langle n(t)\rangle\right].
\end{equation}
We consider equal temperatures, $n_c(t)=n_h(t) = \bar{n}-\frac{\hbar \omega_0}{k_B T}\bar{n}(1+\bar{n})K(t)$ to first order in $K(t)$. Introducing $\delta n(t) = \langle n(t)\rangle - \bar{n}$, we get
\begin{equation}
\frac{d\delta n(t)}{dt} = -(\gamma_c+\gamma_h)\delta n(t)-(\gamma_c+\gamma_h)\frac{\hbar \omega_0}{k_BT}\bar{n}(1+\bar{n})K(t).
\end{equation}
In the Fourier domain this gives
\begin{equation}
\delta n(\omega) = -\frac{\hbar \omega_0}{k_BT}\frac{(\gamma_c+\gamma_h)\bar{n}(1+\bar{n})}{\gamma_c+\gamma_h+i\omega}K(\omega).
\end{equation}
or
\begin{equation}
\Delta U(\omega) =\hbar\omega_0\delta n(\omega)=- \frac{(\hbar \omega_0)^2}{k_BT}\frac{(\gamma_c+\gamma_h)\bar{n}(1+\bar{n})}{\gamma_c+\gamma_h+i\omega}K(\omega).
\end{equation}
From this we find the susceptibility
\begin{equation}
\chi(\omega) = \frac{\Delta U(\omega)}{K(\omega)}= -\frac{(\hbar \omega_0)^2}{k_BT}\frac{(\gamma_c+\gamma_h)\bar{n}(1+\bar{n})}{\gamma_c+\gamma_h+i\omega}.
\end{equation}
In particular, we have
\begin{equation}
\text{Im}[\chi(\omega)] = \frac{(\hbar \omega_0)^2}{k_BT}\bar{n}(1+\bar{n})\frac{(\gamma_c+\gamma_h)\omega}{(\gamma_c+\gamma_h)^2+\omega^2}.
\label{Im}
\end{equation}
Comparing Eqs.~\eqref{S_U expression} and \eqref{Im}, we then find the FDT
\begin{equation}
S_U(\omega) = 2k_B T\frac{\text{Im}[\chi(\omega)]}{\omega},
\end{equation}
or
\begin{equation}
S^c_Q(\omega)+S^h_Q(\omega)+2\textrm{Re}[S^{ch}_Q(\omega)] = 2k_B T \omega \text{Im}[\chi(\omega)],
\end{equation}
which is Eq.~\eqref{Eq14} in the main text.

\end{appendix}

\end{document}